\documentclass[a4paper,11pt]{article}
 \pdfoutput=1 % if your are submitting a pdflatex (i.e. if you have
\usepackage{jheppub}
\usepackage[T1]{fontenc} 
\usepackage{amssymb}
\usepackage{amsmath}
\usepackage{color}
\usepackage{xspace}
\usepackage{caption}
\usepackage{appendix}
\usepackage{float}
\usepackage{hyperref}
\usepackage{diagbox}
\usepackage{soul}
\usepackage{slashed}
\usepackage{amsfonts}
\usepackage{multirow}
\usepackage{mathrsfs}
\usepackage{graphicx}
\usepackage{amsmath}
\usepackage{amssymb}
\usepackage{bm}
\usepackage{bbm}
\usepackage{color}
\usepackage{comment}
\usepackage{subfigure}
\usepackage{textcomp}

\newcommand{\nn}{\nonumber}
\newcommand{\as}{\alpha_s}
\newcommand{\varA}[1]{{\operatorname{#1}}}

\title{Higgs boson pair production and decay at NLO in QCD:  the  $b\bar{b}\gamma\gamma$ final state}

\author{Hai Tao Li,$^a$ Zong-Guo Si,$^a$ Jian Wang,$^{a,b}$ Xiao Zhang,$^a$ Dan Zhao$^a$}

\affiliation{$^a$School of Physics, Shandong University, Jinan, Shandong 250100, China}
\affiliation{$^b$Center for High Energy Physics, Peking University, Beijing 100871, China}

\emailAdd{haitao.li@sdu.edu.cn}
\emailAdd{zgsi@sdu.edu.cn}
\emailAdd{j.wang@sdu.edu.cn}
\emailAdd{zhaangx@mail.sdu.edu.cn}
\emailAdd{zhaoodan@mail.sdu.edu.cn}

\abstract{
The Higgs boson pair production at the LHC provides a probe to the Higgs boson self-coupling.  
The higher-order QCD corrections in this process are sizable and must be taken into account in comparison with data.
Due to the small cross section, it is necessary to consider at least one of the Higgs bosons decaying to bottom quarks.
The QCD corrections to the decay processes would also be important in such cases.
We present a full calculation of the total and differential cross sections for the $b\bar{b}\gamma\gamma$ final state with next-to-leading order (NLO) QCD corrections.
After applying typical kinematic cuts in the final state,
we find that QCD NLO corrections in the decay decrease the LO result by $19\%$ and reduce the scale uncertainties by a factor of two.
The QCD corrections to the invariant mass $m_{jj\gamma\gamma}$ distribution, 
the transverse momentum spectra of the leading bottom quark jet and photon are significant and can not be approximated by a constant factor.
}

\begin{document} 
\maketitle
\flushbottom

\section{Introduction}
\label{sec:intro}

The discovery of the Higgs boson more than ten years ago \cite{ATLAS:2012yve, CMS:2012qbp} marks the great success of the standard model (SM), and inaugurates a new era in particle physics.
The presence of a Higgs field is essential to explain the masses of the $W$ and $Z$ gauge bosons via the Higgs mechanism \cite{PhysRevLett.13.321, HIGGS1964132, PhysRevLett.13.508, PhysRevLett.13.585}.
Until now, the properties of this Higgs boson, including  its mass and width \cite{CMS:2020xrn,ATLAS:2023zhs,CMS:2022ley}, spin and parity \cite{ATLAS:2015zhl,CMS:2014nkk}, as well
as various production and decay rates \cite{ATLAS:2022vkf,CMS:2022dwd},  have been measured, 
and the results are in agreement with the SM predictions.

A precise understanding of the trilinear Higgs self-coupling $\lambda_{HHH}$ 
is crucial for unraveling the mechanism of electroweak symmetry breaking and offering insights into vacuum stability \cite{Degrassi:2012ry}.
The self-interaction of the Higgs boson contributes to higher-order electroweak corrections in both the production and decay processes of a single Higgs boson, and thus the Higgs self-coupling gets constrained by precisely measuring the cross sections of the single Higgs processes \cite{McCullough:2013rea,Gorbahn:2016uoy,Degrassi:2016wml,Bizon:2016wgr,ATLAS:2022jtk,Gao:2023bll}.
However, this method relies on the assumption that there is no modification of the other Higgs couplings \cite{DiVita:2017eyz}, and the dependence appears as a loop effect.
Therefore, it is more informative to measure the Higgs boson pair productions, 
which are sensitive to the trilinear Higgs self-coupling at leading order (LO).

At the LHC, the gluon fusion is the dominant production channel, in which the Higgs bosons are emitted from a top quark loop.
The cross section of this process has been calculated up to QCD ${\rm N^3LO}$ in the infinite large top quark mass limit \cite{Dawson:1998py,deFlorian:2013jea,deFlorian:2016uhr,Chen:2019lzz,Chen:2019fhs} and up to QCD NLO with finite $m_t$ dependence \cite{Borowka:2016ehy, Borowka:2016ypz, Baglio:2018lrj, Baglio:2020ini}.
In the $m_t\to \infty$ limit,
the NLO corrections are significant, increasing the LO results by $87\%$,
and the NNLO corrections provide an additional $33\%$ enhancement. 
The ${\rm N^3LO}$ corrections amount to about $6.4\%$ of the LO cross sections.
With finite top quark mass, the NLO QCD corrections increase the LO cross section by $66\%$.
Partial $m_t$ effects at NNLO in QCD and the electroweak corrections at NLO have been studied in \cite{Grazzini:2018bsd,Czakon:2020vql,Mazzitelli:2022scc,Davies:2023obx} and \cite{Borowka:2018pxx,Muhlleitner:2022ijf,Davies:2022ram,Davies:2023npk,Bi:2023bnq}, respectively.
Soft gluon resummation effects have been taken into account in \cite{Shao:2013bz,deFlorian:2015moa,Ajjath:2022kpv}.

The CMS collaboration has searched for the Higgs pair productions in the $b\overline{b}\gamma\gamma$, $b\overline{b}b\overline{b}$, $b\overline{b}\tau\tau$, $b\overline{b}ZZ$ and multilepton final states, giving a limit of 3.4 times the cross section predicted by the SM at $95\%$ confidence level (CL),
which is converted to the allowed range for the self-coupling $-1.24<\kappa_{\lambda}(=\lambda_{HHH}/\lambda_{HHH}^{\rm SM})<6.49$ at $95\%$ CL \cite{CMS:2022dwd}.
The ATLAS collaboration has measured the Higgs pair productions in the $b\overline{b}\gamma\gamma$, $b\overline{b}\tau\tau$, and $b\overline{b}b\overline{b}$ final states, 
and set an upper limit of 2.4 times SM predictions.
Combined with the single Higgs boson analyses, this is transformed to the constraint $-0.4<\kappa_{\lambda}<6.3$ at $95\%$ CL \cite{ATLAS:2022jtk}.
The  range would be significantly narrowed at the high-luminosity LHC \cite{ATLAS:2022qjq}.

In obtaining the above limits, precise predictions for the kinematic distributions of the $HH$ events, especially the $m_{HH}$ distribution, play an important role \cite{ATLAS:2019qdc}.
So far, only the higher-order QCD corrections in the production processes have been studied.
It is not clear whether the kinematic distributions would change after considering the QCD corrections in the decays of the Higgs bosons.
Our aim in this paper is to address this issue.
Because the cross section of the Higgs boson pair production is already very small, it is required to search for the signal events with at least one of the Higgs bosons decaying to two bottom quarks.
We focus on the $ HH\to b\bar{b}\gamma\gamma$ final state in the present work.
We find that the NLO QCD correction in the decay could reduce the cross section for the process $pp\to HH\to b\bar{b}\gamma\gamma$ significantly and that
the impact on the kinematic distributions can not be described by a constant.

This paper is organized as follows.
In section \ref{sec:calcframe}, we introduce the calculation framework.
The numerical results for the total cross sections and kinematical distributions are presented in section \ref{sec:numresult}.
The conclusion is given in section \ref{sec:conclusion}.

\section{Calculation framework}
\label{sec:calcframe}

We are going to calculate the full QCD corrections to $gg\to HH$ production with the $HH\to b\bar{b}\gamma\gamma$ decay at the LHC.
The interference between the QCD corrections in the production and decay processes starts from NNLO 
and is suppressed by a factor of $\Gamma_H/m_H\sim 10^{-5}$ compared to the resonance contribution, in which the Higgs bosons are taken to be on-shell.
In this paper, we consider only the $\mathcal{O}(\alpha_s)$ correction and therefore neglect the interference.

In the narrow-width approximation, the cross section is written as
\begin{align}
\sigma_{\rm pro+dec} = \sigma_{\rm pro} \frac{1}{\Gamma_{H_1}} \frac{1}{\Gamma_{H_2}} \Gamma_{\rm dec}
\end{align}
where $ \sigma_{\rm pro}$ denotes the production cross section of $gg\to H_1 H_2$ and $\Gamma_{H_1}=\Gamma_{H_2}=\Gamma_{H}$ is the total width of the Higgs boson. 
The double decay  width is defined by
\begin{align}
   \Gamma_{\rm dec} \equiv \int d \Gamma_{H_1}    \int d \Gamma_{H_2} ~ F_J
\end{align}
with $F_J$ the measurement function which embodies the cut information in event selection.
For a specific decay mode, i.e., 
$H_1 \to X_1$ and $H_2\to X_2$, 
the cross section is expressed as
\begin{align}
\sigma_{\rm pro+dec(X_1,X_2)} = \sigma_{\rm pro} \frac{1}{\Gamma_{H_1\to X_1}} \frac{1}{\Gamma_{H_2\to X_2}} \Gamma_{\rm dec(X_1,X_2)} \times R(H_1\to X_1) R(H_2\to X_2)
\end{align}
where we have introduced $R(H_i\to X_i)$ as the branching ratio for the Higgs decay into the final state $X_i$.

We focus on the higher-order QCD corrections to the production and decay to a specific final state $X_1+X_2$.
The cross sections and decay widths can be expanded in a series of the strong coupling constant $\alpha_s$, e.g., 
\begin{align}
    \sigma = \sigma^{(0)} + \sum_{n=1} \alpha_s^n \sigma^{(n)}.
\end{align}
And therefore, we have
\begin{align}
\sigma_{\rm pro+dec(X_1,X_2)}^{(n)} & = \sigma_{\rm pro} \frac{1}{\Gamma_{H_1\to X_1}} \frac{1}{\Gamma_{H_2\to X_2}} \Gamma_{\rm dec(X_1,X_2)}\bigg|_{{\rm expanded~to}~\alpha_s^n}\nn\\
&
\times R(H_1\to X_1) R(H_2\to X_2)\,.
\end{align}
We have taken the branching ratios as constants in our calculation,
and the most precise values given in refs. \cite{LHCHiggsCrossSectionWorkingGroup:2016ypw,Djouadi:2018xqq,Djouadi:1997yw} are adopted. 

As a first step, we incorporate only the NLO QCD corrections in this study.
The  $\mathcal{O}(\as)$ correction is decomposed into two parts,
\begin{align}
\sigma_{\rm pro+dec(b\bar{b},\gamma\gamma)}^{(1)} & = 
\sigma_{\rm pro+dec(b\bar{b},\gamma\gamma)}^{\rm pro(1)}
+\sigma_{\rm pro+dec(b\bar{b},\gamma\gamma)}^{\rm dec(1)}
\label{eq:master}
\end{align}
with
\begin{align}
\sigma_{\rm pro+dec(b\bar{b},\gamma\gamma)}^{\rm pro(1)}=\sigma_{\rm pro}^{(1)} \frac{1}{\Gamma_{H_1\to b\bar{b}}^{(0)}} \frac{1}{\Gamma_{H_2\to \gamma\gamma}^{(0)}} \Gamma^{(0)}_{\rm dec(b\bar{b},\gamma\gamma)}
\times R(H_1\to b\bar{b}) R(H_2\to \gamma\gamma)
\label{eq:masterpro}
\end{align}
and
\begin{align}
\sigma_{\rm pro+dec(b\bar{b},\gamma\gamma)}^{\rm dec(1)}&=
\sigma_{\rm pro}^{(0)} \frac{1}{\Gamma_{H_1\to b\bar{b}}^{(0)}} \frac{1}{\Gamma_{H_2\to \gamma\gamma}^{(0)}} \Gamma^{(1)}_{\rm dec(b\bar{b},\gamma\gamma)}
\times R(H_1\to b\bar{b}) R(H_2\to \gamma\gamma)\nn\\
&-
\sigma_{\rm pro}^{(0)} \frac{\Gamma_{H_1\to b\bar{b}}^{(1)}}{\left( \Gamma_{H_1\to b\bar{b}}^{(0)}\right)^2} \frac{1}{\Gamma_{H_2\to \gamma\gamma}^{(0)}} \Gamma^{(0)}_{\rm dec(b\bar{b},\gamma\gamma)}
\times R(H_1\to b\bar{b}) R(H_2\to \gamma\gamma) \nn \\
&-
\sigma_{\rm pro}^{(0)} \frac{1}{ \Gamma_{H_1\to b\bar{b}}^{(0)}} \frac{\Gamma_{H_2\to \gamma\gamma}^{(1)}}{\left(\Gamma_{H_2\to \gamma\gamma}^{(0)}\right)^2} \Gamma^{(0)}_{\rm dec(b\bar{b},\gamma\gamma)}
\times R(H_1\to b\bar{b}) R(H_2\to \gamma\gamma)\,.
\label{eq:decaynlo}
\end{align}
If no cuts are applied and the full phase space of the final state particles is integrated over, 
$\sigma_{\rm pro+dec(b\bar{b},\gamma\gamma)}^{\rm dec(1)}$ is vanishing.
In practice, kinematic cuts have to be imposed to select the events. 
In this case, $\sigma_{\rm pro+dec(b\bar{b},\gamma\gamma)}^{\rm dec(1)}$ may provide a significant correction.
It is our purpose to investigate the contribution of this part.

Because there is no real correction from $H\to \gamma\gamma +g$ at NLO in QCD due to color conservation, the decay $H\to \gamma\gamma$ receives only virtual corrections.
Then the third line in eq.(\ref{eq:decaynlo}) could cancel with part of the first line that is given by
\begin{align}
    \int d \Gamma_{H_1\to b\bar{b}}^{(0)} \times   \int d \Gamma_{H_2\to \gamma\gamma}^{(1)}
    \label{eq:dhrr}
\end{align}
since they have the same kinematics.
As a consequence, 
\begin{align}
\sigma_{\rm pro+dec(b\bar{b},\gamma\gamma)}^{\rm dec(1)}&=
 \sigma_{\rm pro}^{(0)} \frac{1}{\Gamma_{H_1\to b\bar{b}}^{(0)}} \frac{1}{\Gamma_{H_2\to \gamma\gamma}^{(0)}} \Gamma^{(0)}_{\rm dec(b\bar{b},\gamma\gamma)}
\left( \frac{\int d \Gamma_{H_1\to b\bar{b}}^{(1)} F_J }{  \int d \Gamma_{H_1\to b\bar{b}}^{(0)} F_J } -  \frac{\Gamma_{H_1\to b\bar{b}}^{(1)}}{ \Gamma_{H_1\to b\bar{b}}^{(0)}}\right)
\nn\\
&
\times R(H_1\to b\bar{b}) R(H_2\to \gamma\gamma) \,.
\label{eq:masterdec}
\end{align}
We have written the first term in the bracket in differential form because it is subject to the kinematic cuts.

The above eqs.(\ref{eq:master},\ref{eq:masterpro},\ref{eq:masterdec}) represent the master formulae of our calculation.
The $\mathcal{O}(\as)$ correction to the production process $\sigma_{\rm pro}^{(1)}$ has been computed in \cite{Borowka:2016ypz} with full top quark mass dependence.
We have checked the results independently using our code, although the two-loop $gg\to HH$ amplitude is calculated via the grid implemented in POWHEG BOX \cite{Alioli:2010xd}.
The one-loop $gg\to HHg$ and other similar $2\to 3$ partonic amplitudes are generated by the automatic tool OpenLoops \cite{Buccioni:2019sur,Buccioni:2017yxi} with the scalar integrals evaluated by Collier/OneLoop \cite{Denner:2016kdg,vanHameren:2010cp}.
All the tree-level and one-loop amplitudes of the decay process $H\to b\bar{b}+X$ are calculated employing the FeynArts \cite{KUBLBECK1990165,Hahn:2000kx} and FeynCalc \cite{MERTIG1991345,Shtabovenko:2016sxi,Shtabovenko:2020gxv} packages.
The analytical results are collected in Appendix \ref{sec:Hbb}.
The final-state bottom quarks have been taken to be massless while
the Yukawa coupling of the bottom quark is kept finite.
The result of the squared amplitude for the decay $ H\to \gamma\gamma $  is recorded in Appendix \ref{sec:Haa}.
The infrared divergences in the virtual and real corrections are subtracted using the dipole method \cite{Catani:1996vz,Gleisberg:2007md}.
In constructing the analytical subtraction terms, we need the one-loop amplitude of $gg\to HH$ which is evaluated with the aid of QCDLoop \cite{Carrazza:2016gav}.

\section{Numerical results}
\label{sec:numresult}

In numerical calculation, the SM input parameters are set to be 
\begin{align}
& G_F = 1.1663787\times10^{-5} ~{\rm GeV}^{-2} ,
\quad m_W=80.399 ~{\rm GeV}, \nn \\
& m_H = 125 ~{\rm GeV}, \quad 
m_t = 173 ~{\rm GeV}, \quad
m_b(m_b) =4.18 ~{\rm GeV},\nn\\
& R(H\rightarrow bb)=0.5824, \quad  R(H\rightarrow \gamma\gamma)=2.27\times10^{-3} .
\end{align}
We choose the parton distribution function (PDF) set PDF4LHC15\_nlo\_100\_pdfas \cite{Buckley:2014ana}.
The strong coupling is evaluated using the function associated with this PDF set. 
The default values of the factorization scale $\mu_{\rm F}$ 
and renormalization scale $\mu_{\rm R}$ are set as $m_{HH}/2$.
In the calculation of the real corrections to $pp\to HH$, we have applied a phase space cut $p_T^{\rm cut} \sim 0.1 $ GeV on the Higgs pair system to ensure numerical stability. 
Below this cut, the contributions from the squared amplitude and the differential subtraction terms cancel against each other and could be neglected.
We have checked that the total cross sections are insensitive to this cut when it is chosen from 0.001 GeV to 1 GeV.
A kinematic cut $s_{ij}^{\rm cut}\sim 10^{-5} \, {\rm GeV^2}$ is applied to the final-state partons when considering the higher-order QCD corrections to $H\to b\bar{b}$ due to the same reason. 

\begin{figure}
    \centering
    \includegraphics[width=0.46\linewidth]{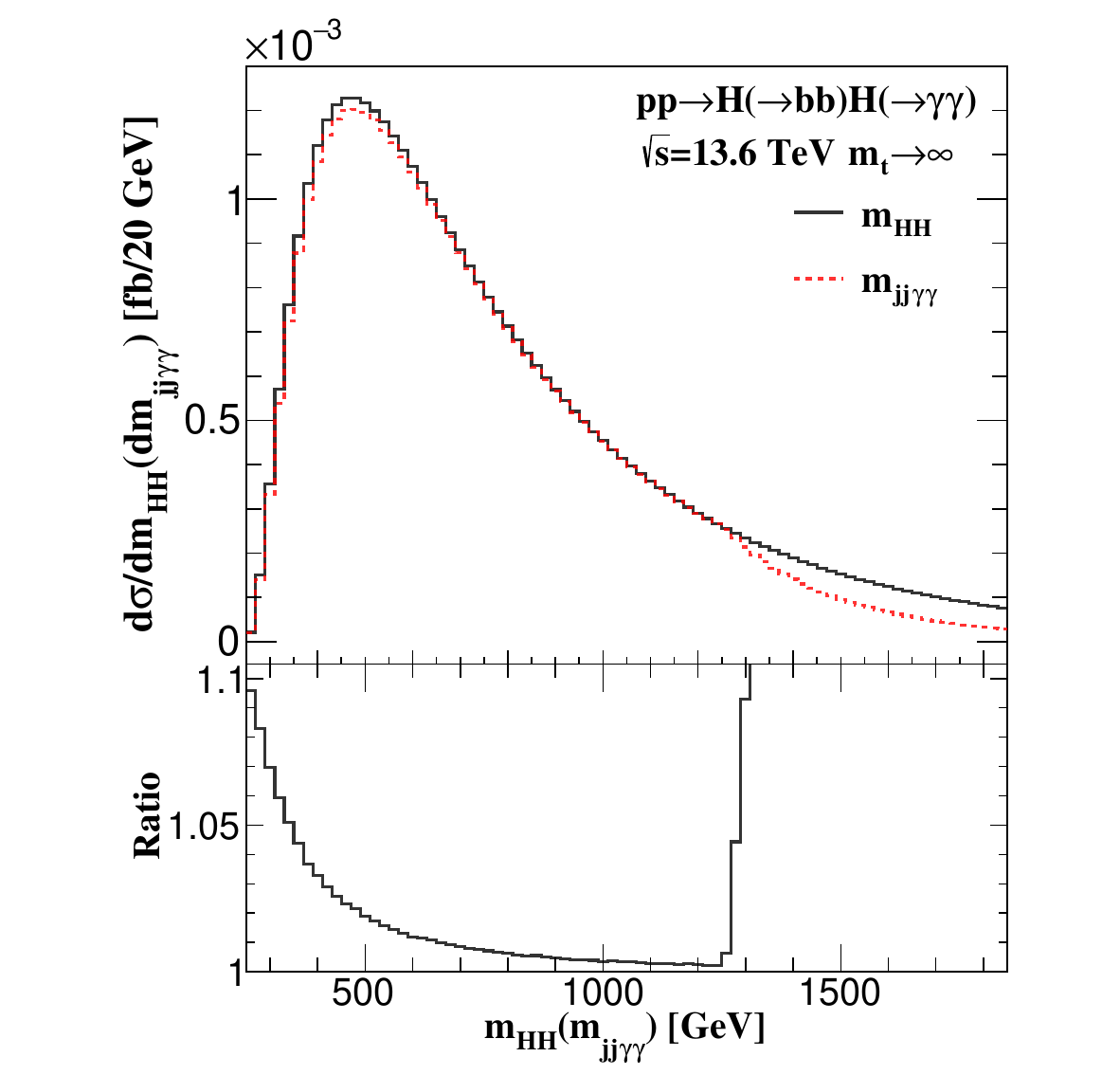}
    \includegraphics[width=0.46\linewidth]{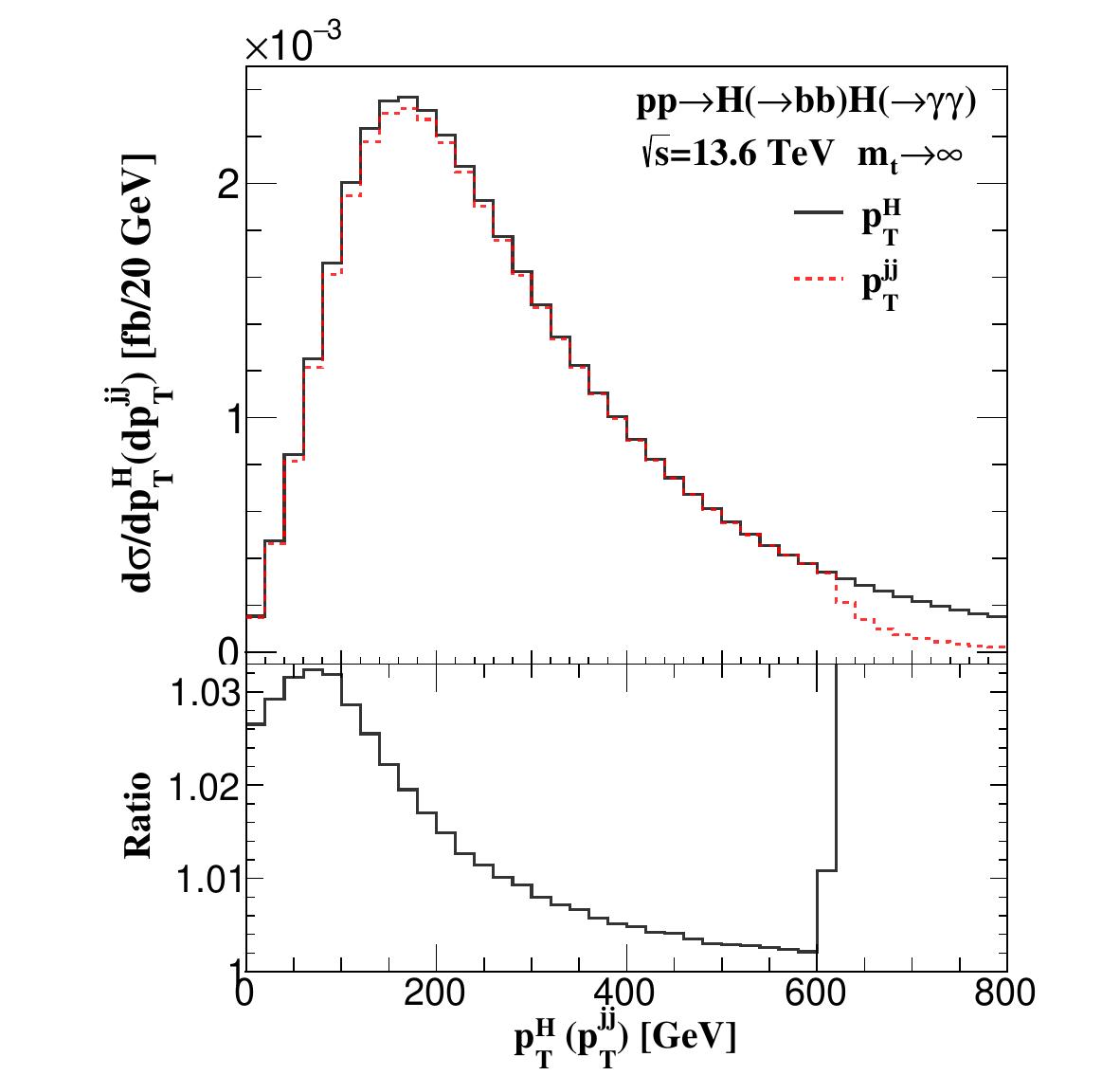}    
    \caption{Comparison of the kinematic distributions for the intermediate and reconstructed Higgs bosons in the heavy top quark mass limit at LO. The lower panels show the ratio of the distributions for the intermediate Higgs bosons over those for the reconstructed Higgs bosons. The ratio rises sharply for $m_{jj\gamma\gamma}$ larger than 1200 GeV and can not be shown properly in the plot. The value is about 2.5 at $m_{jj\gamma\gamma}=1800$ GeV. In the right plot, the ratio is about 7 at  $p_T^{jj}=$ 800 GeV. }
    \label{fig:comparisonHiggs}
\end{figure}

Before we show any numerical results, it is necessary to elaborate on the importance 
of including decay processes in the theoretical calculation.
One crucial reason is that all the events in experiments are collected under kinematic cuts.
Therefore it is the cross section with cuts that is measured.
In experiments, the cuts can only be applied to the stable particles, i.e., the jets, photons, and leptons.
One has to take into account the decay processes of unstable particles to fully simulate the events.

For illustration, 
we compare the kinematic distributions of the intermediate Higgs boson resonances and the reconstructed Higgs bosons from the final-state $b$-jets and photons in figure \ref{fig:comparisonHiggs}.
The intermediate Higgs boson appears in the final state of the production process but decays to bottom quarks or photons immediately. 
Its momentum is known in the theoretical calculation but can not be measured directly.
In contrast, the momentum of the reconstructed Higgs boson is defined as the sum of the momenta of the two $b$-jets, denoted by $j$, or the two photons.
The final-state (anti-)bottom quarks and other partons are clustered to jets using the anti-$k_t$ algorithm \cite{Cacciari:2008gp,Cacciari:2011ma} with a separation parameter $R=0.4$.
The angular distance between the photons $\gamma$ and the $b$-jets $j$ should be larger than 0.4.
From figure     \ref{fig:comparisonHiggs}, we see that the reconstructed invariant mass $m_{jj\gamma\gamma}$ spectrum is generally lower than the intermediate Higgs boson invariant mass $m_{HH}$ spectrum by about $1\%\sim 10\%$ at $m_{HH} (m_{jj\gamma\gamma}) < 600$ GeV.
This is mainly because the phase space with $R_{j\gamma}<R$ has been cut out.
The difference becomes smaller with the increasing of $m_{HH} (m_{jj\gamma\gamma})$,
because the $b$-jets and photons tend to fly away from each other for larger  $m_{HH} (m_{jj\gamma\gamma})$.
The single Higgs transverse momentum distribution is lowered by about $2\%\sim 3\%$ in the reconstructed kinematics for $p_T^{jj}< 200$ GeV,
and the deviation between the distributions using the reconstructed and intermediate Higgs boson becomes smaller as $p_T^{jj} (p_T^H)$ increases from 100 GeV to 600 GeV due to the same reason in the discussion of $m_{HH}$. 

An obvious deviation can be found in the region of $m_{HH}>1250$ GeV or $p_T^H> 620$ GeV,
where the reconstructed distribution drops quickly.
The reason is that the bottom quark pair from the Higgs decay is highly boosted so that they become collinear with each other 
and are subject to the requirement  $R_{jj}>R$.
The threshold value of $p_T^H$ can be estimated by
\begin{align}
    \frac{2m_H}{R}
\end{align}
which is the minimum of the momentum $p_H=m_H/R/\sqrt{z(1-z)}$  of the Higgs boson decaying to a collinear bottom (anti-)quark pair with a separation angle of $R$ and the momentum fraction $z$ carried by one bottom quark.
One can not observe such an interesting feature without considering the decays of the Higgs bosons.
Note that the distributions presented in figure \ref{fig:comparisonHiggs} are obtained in the heavy top quark mass limit to show the features clearly. The distributions with finite top quark mass fall off fast. The ratios, however, exhibit similar shapes.  

In practice, the particles with very low transverse momentum or very large rapidity are not detected.
To calculate cross sections with decays, we apply the following cuts that are similar to those used 
in the analysis of the CMS collaboration \cite{CMS:2020tkr}, 
\begin{align}
   & p_T^j \ge 25 ~{\rm GeV}, \quad
    p_T^{\gamma} \ge 25 ~{\rm GeV}, \quad
    |\eta^j| \le 2.5, \quad
    |\eta^{\gamma}| \le 2.5, \quad \nn \\
&  90~ {\rm GeV} \le m_{jj} \le 190 ~{\rm GeV}, \quad R_{jj,j\gamma,\gamma\gamma}\ge 0.4.
\label{eq:cuts}
\end{align}
These cuts are rather loose because we want to keep more signals after cuts.

\begin{table}[ht]
	\centering
	\begin{tabular}{|c|c|c|c|c|c|}
		\hline
		&without decays & \multicolumn{2}{c|}{with decays but no cuts}& \multicolumn{2}{c|}{with decays and cuts}\\ \cline{1-6}
		& & ${\rm LO}^{\rm dec}$ & $\delta {\rm NLO}^{\rm dec}$ & ${\rm LO}^{\rm dec}$ & $\delta {\rm NLO}^{\rm dec} $  \\ 
  \cline{1-6}
		${\rm LO}_\infty^{\rm pro}$ &$17.07_{-22\%}^{+31\%}$ &$0.04514_{-22\%}^{+31\%}$ &$0$&$0.02515^{+30\%}_{-22\%}$ &$-0.00349^{+42\%}_{-28\%}$ \\ \cline{1-1}
		${\rm LO}_{m_t}^{\rm pro}$ &$19.85^{+28\%}_{-20\%}$& $0.05248^{+28\%}_{-20\%}$&$0$&$0.02789^{+27\%}_{-20\%}$&$-0.00523^{+39\%}_{-27\%}$ \\ \cline{1-1}
		$\delta {\rm NLO}_{\infty}^{\rm pro}$ & $14.86_{-7\%}^{+6\%}$&$0.03928^{+6\%}_{-7\%}$ & $-$&$0.02128^{+6\%}_{-7\%}$&$-$\\ \cline{1-1}
		$\delta {\rm NLO}_{m_t}^{\rm pro}$ & $13.08^{+4\%}_{-8\%}$ &$0.03458^{+4\%}_{-8\%}$ &$-$ &$0.01829^{+4\%}_{-8\%}$&$-$\\ \cline{1-6} 
            \multicolumn{6}{|c|}{Full NLO result} \\  \cline{1-6}
            ${\rm NLO}_{\infty}$ & $31.93^{+18\%}_{-15\%}$ & \multicolumn{2}{c|}{$0.08442^{+18\%}_{-15\%}$}& \multicolumn{2}{c|}{$0.04294^{+15\%}_{-14\%}$}\\ \cline{1-1}
            ${\rm NLO}_{m_t}$ & $32.93^{+14\%}_{-13\%}$ & \multicolumn{2}{c|}{$0.08706^{+14\%}_{-13\%}$}& \multicolumn{2}{c|}{$0.04095^{+10\%}_{-11\%}$}\\ \cline{1-6} 
	\end{tabular}
 \caption{Total cross sections (in fb) of the Higgs boson pair production and decays $gg\to HH\to b\bar{b}\gamma\gamma$ at the 14 TeV LHC. 
 %The PDF set is ${\rm PDF4LHC15\_nlo\_100\_pdfas}$. 
 The relative uncertainties are obtained by taking the seven-point scale variations around the default values.
 The subscripts $m_t$ and $\infty$ indicate that the results are calculated with full $m_t$ dependence or in the large $m_t$ limit, respectively. The notation $\delta {\rm NLO}$ denotes the $\mathcal{O}(\as)$ correction while the NLO results include both LO and  $\delta {\rm NLO}$ contributions. 
 The symbol `$-$' represents the higher-order corrections that we neglected in this work. 
 The results without decays have been multiplied by $1/2$ to follow the convention of identical Higgs boson pair production.}
  \label{tab:totalxs}
\end{table}

In table \ref{tab:totalxs} we show the total cross sections with or without decays or cuts.
The second column gives the total cross section without decays.
We have performed the calculations up to NLO in QCD
both in the infinitely large $m_t$ limit and with full $m_t$ dependence.
Our results coincide with the previous computations 
taken with the infinite $m_t$  \cite{Chen:2019fhs} and finite $m_t$ \cite{Borowka:2016ypz} setup if the same PDF sets and SM input parameters are adopted.  
The LO production cross section with infinite $m_t$ is smaller than that with finite $m_t$ by $14\%$. 
But it receives larger QCD corrections, e.g., $87\%$ in the infinite $m_t$ limit compared to $66\%$ with full $m_t$ dependence.
As a result, the difference between the two schemes is only $3\%$ at NLO. 
We have estimated the scale uncertainty by varying the factorization $\mu_{\rm F}$ and renormalization $\mu_{\rm R}$ scales independently by a factor of two, excluding the cases when the ratio of the two scales $\mu_{\rm F}/\mu_{\rm R}=4,1/4$.
We see that the scale uncertainty is reduced almost by a factor of two after including the QCD corrections.

The third column shows the total cross sections with decays but without cuts.
The results with LO decays agree with the production cross sections multiplied by a product of the branching ratios, which is about $0.27\%$. 
The cross sections with the LO production and  $\mathcal{O}(\as)$ decays are vanishing, 
as expected from eq. (\ref{eq:masterdec}).
These features serve as strong checks of the validity of our numerical program.

The cross sections with kinematics cuts (\ref{eq:cuts}) on the final-state jets and photons are listed in the fourth column.
The results of the LO production and LO decay after the cuts are $56\%$ and $53\%$ of those without cuts in the infinite and finite $m_t$ cases, respectively.
The NLO corrections in the production suffer from similar suppression.
The NLO corrections in the decay decrease the LO cross section by $19\%$ ($14\%$) for finite (infinite) $m_t$.
This effect is much larger than the ${\rm N^3LO}$ QCD correction,
which is about $+6\%$ \cite{Chen:2019fhs}.

\begin{figure}[h]
    \centering
{\includegraphics[width=0.53\linewidth]{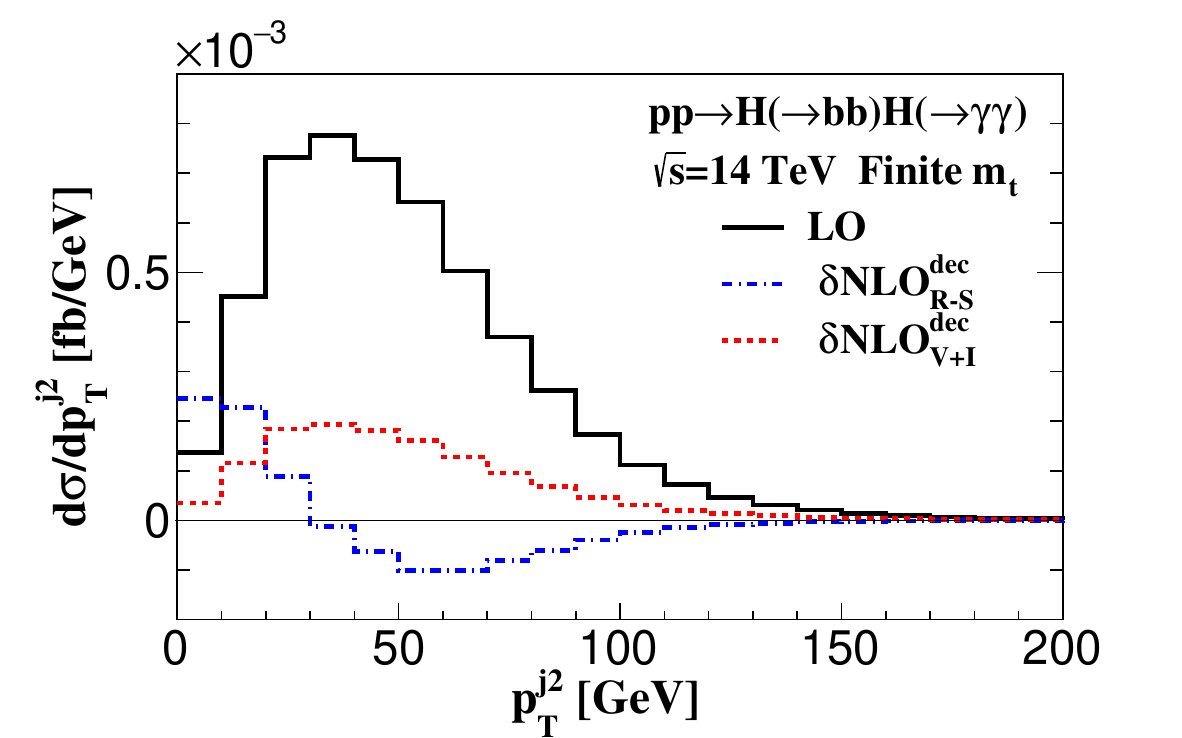}}
   \vspace{0cm}
    \caption{The transverse momentum distribution of the subleading $b$-jet. The black line represents the LO result. The red and blue lines denote the NLO virtual and real corrections with corresponding subtraction terms, respectively.  }
\label{fig:decay_reason}
\end{figure}

This cut rate and the NLO corrections in the decay can be understood by inspecting the transverse momentum distribution for  the subleading $b$-jet, 
which is shown in figure \ref{fig:decay_reason}.
As indicated by the LO result, the transverse momentum spectrum has a peak around 40 GeV.
Therefore a large portion of the events would be neglected under a typical cut $p_T^j\ge 25$ GeV on the jet.
The NLO corrections can be divided into two parts, i.e., the virtual and real corrections  with associated subtraction terms, denoted by $\delta {\rm NLO}_{\rm V+I}$ and $\delta {\rm NLO}_{\rm R-S}$ in the figure, respectively.
The former has the same distribution as the LO because they are proportional to each other with a factor independent of the kinematics; see eq. (\ref{eq:hbbvI}) in the appendix.
However, the latter gives a positive and negative contribution for $p_T<30$ GeV and $p_T>30$ GeV, respectively.
The requirement of a cut $p_T^j>25$ GeV leads to a significant cancellation between the virtual and the real corrections. 
As a result, the first term in the bracket in eq. (\ref{eq:masterdec}) is small and the dominant contribution comes from the second term there, which is around $-20\%$.

\begin{figure}[h]
    \centering
    \subfigure[]{\includegraphics[width=0.49\linewidth]{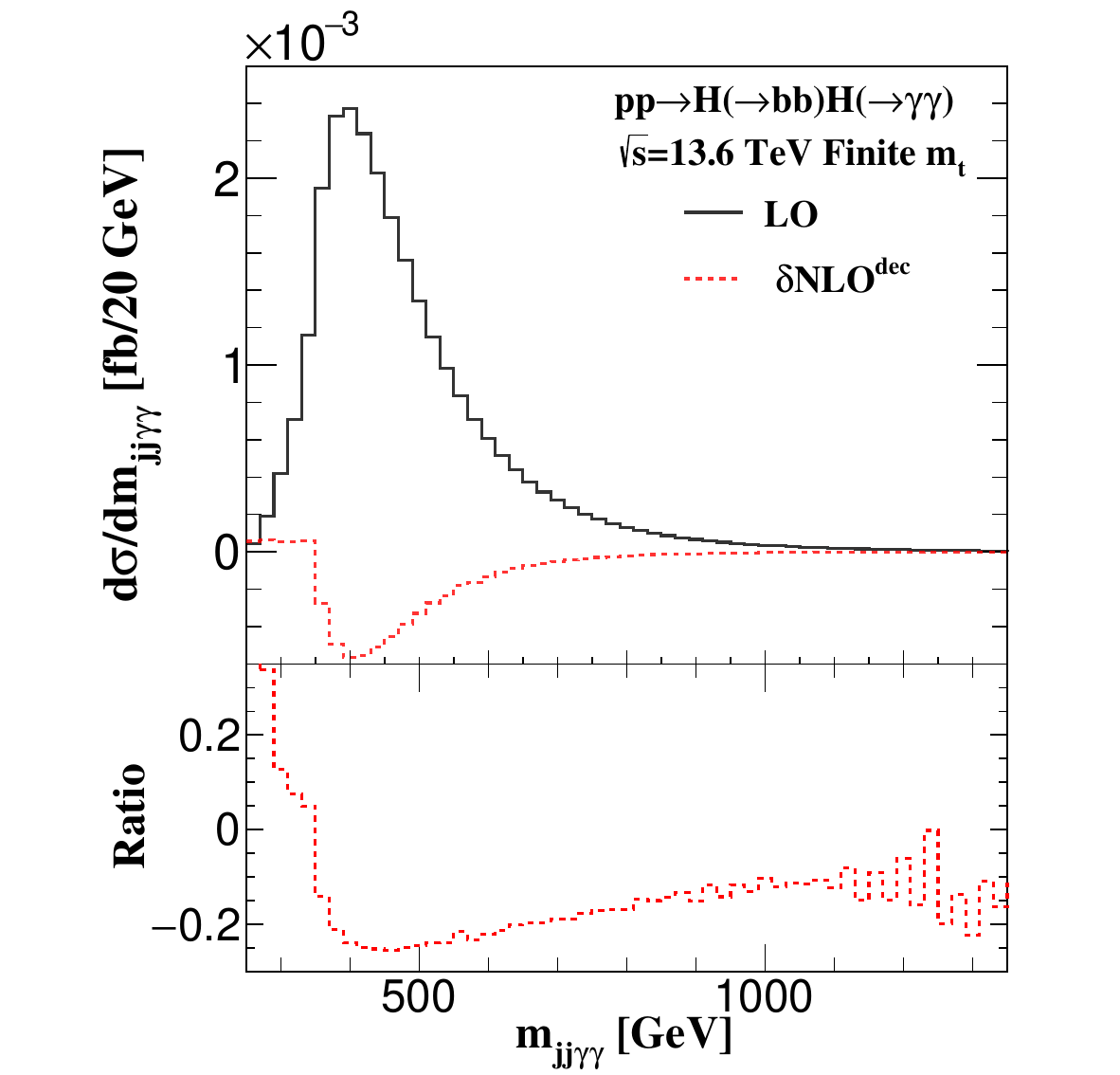}}
   \subfigure[]{\includegraphics[width=0.49\linewidth]{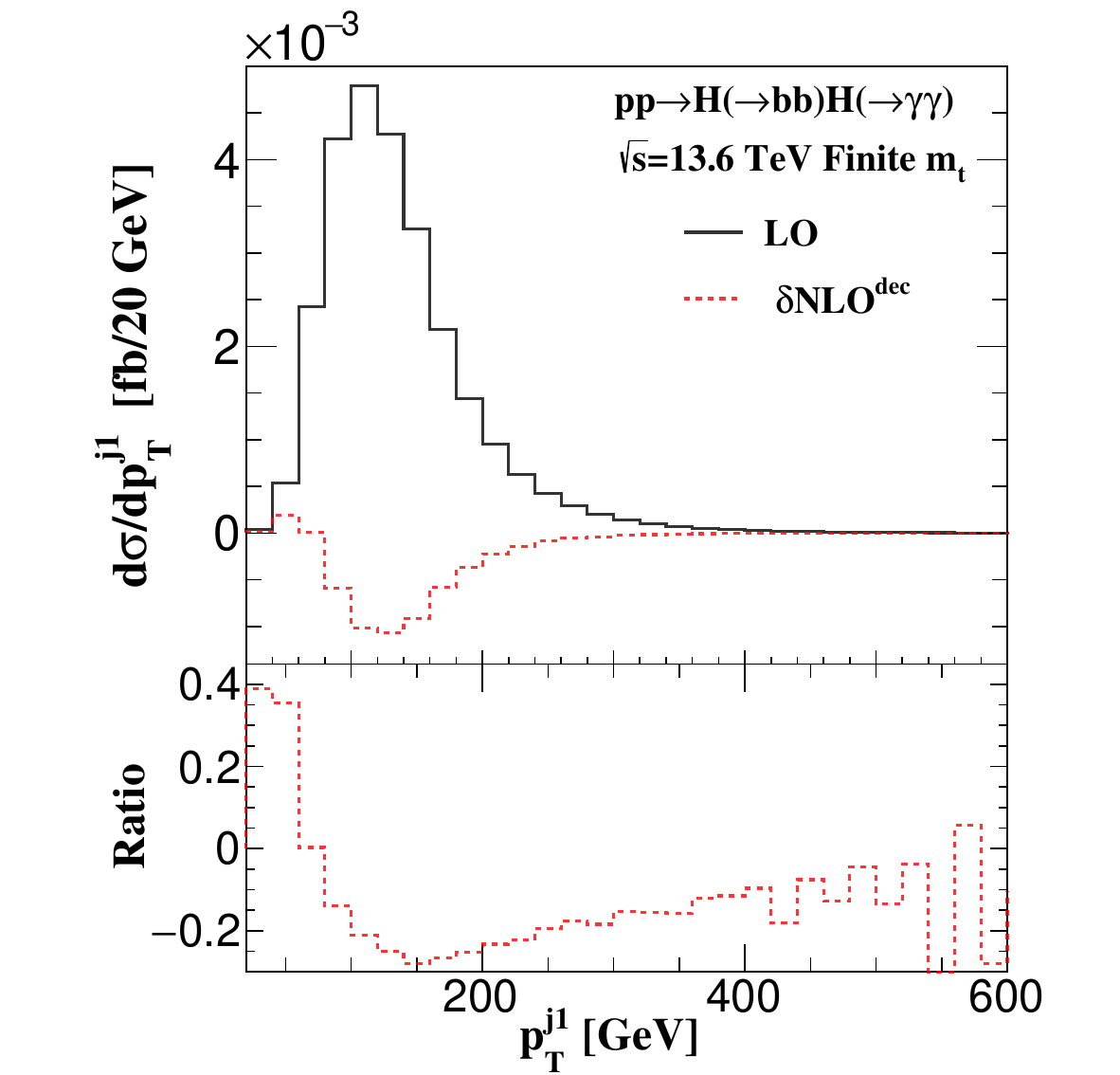}}
   \subfigure[]{\includegraphics[width=0.49\linewidth]{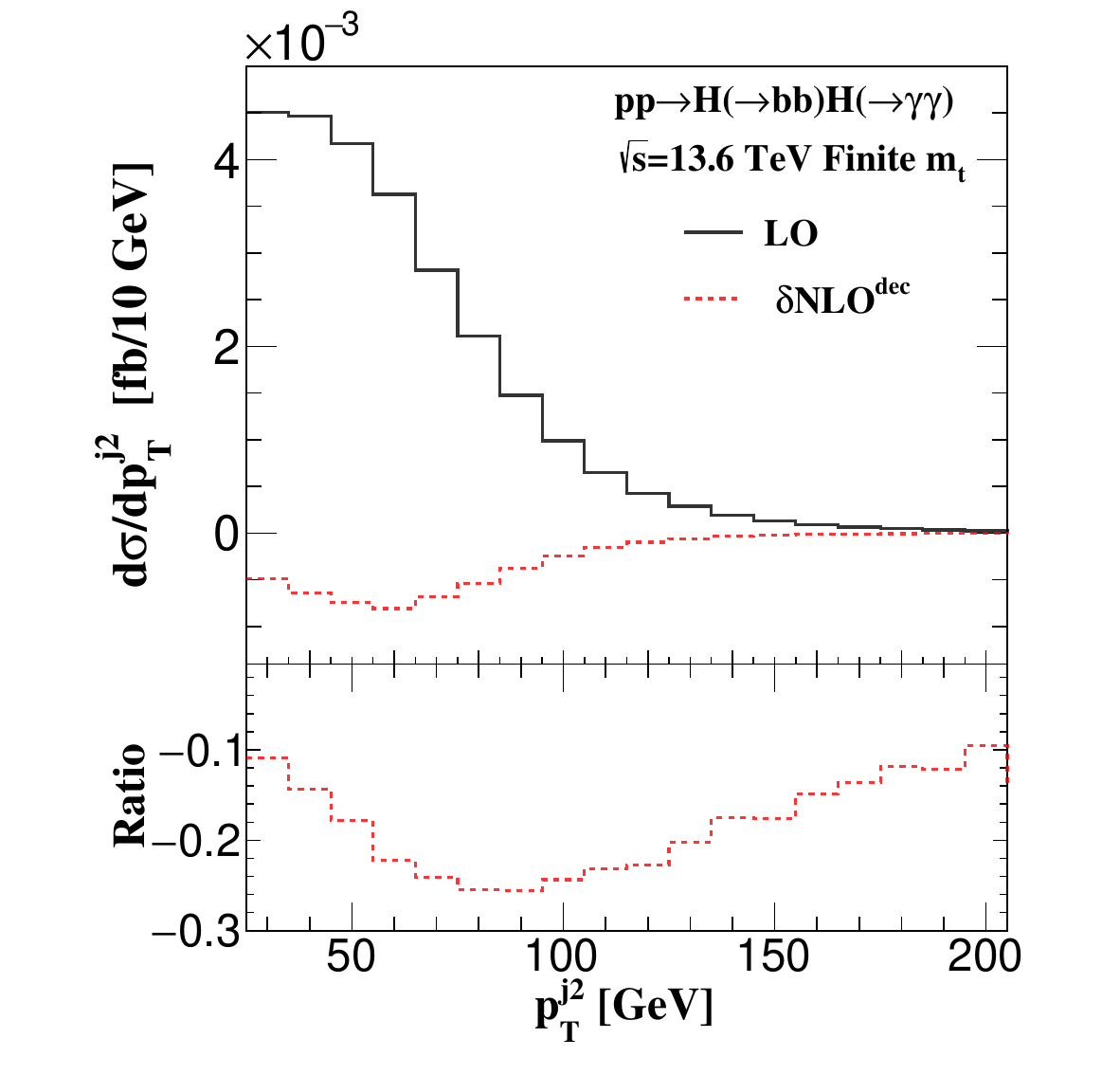}}
   \subfigure[]{\includegraphics[width=0.49\linewidth]{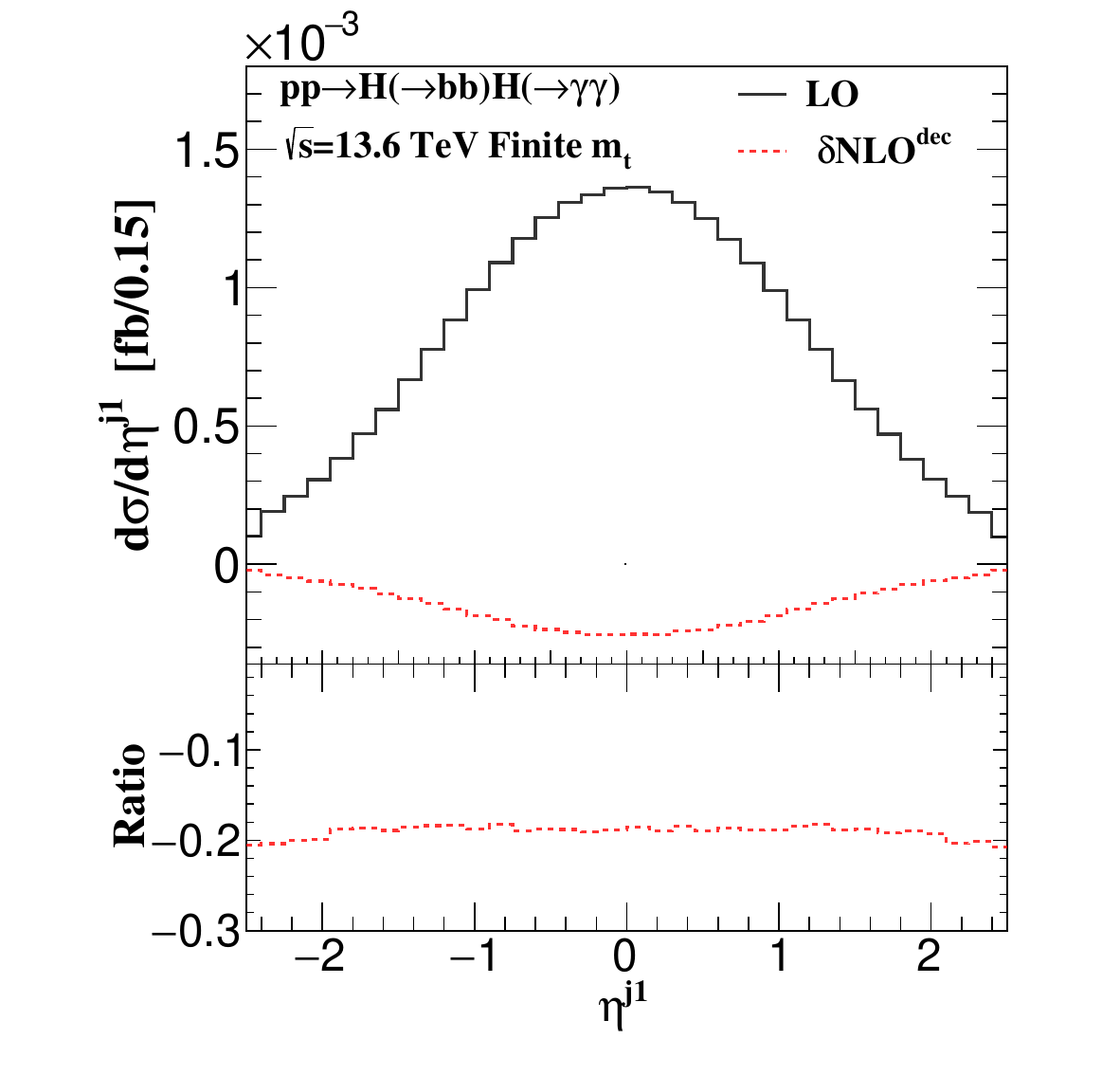}}
   \vspace{0cm}

    \caption{Impact of the QCD NLO corrections in the $H\to b\bar{b}$ decay process on the kinematic distributions  after the cuts (\ref{eq:cuts}). The production cross section is calculated with finite $m_t$.
    The plot $(a)$ shows the invariant mass of the two $b$-jets and two photons.
    The plots $(b)$ and $(c)$ present the transverse momenta of the leading and subleading $b$-jets, respectively. The rapidity distribution of the leading $b$-jet is shown in the plot $(d)$. The ratio in the lower panel in each plot illustrates the magnitude of $\mathcal{O}(\as)$ corrections in the decay process compared to the LO result.}
   \label{fig:decay_NLO_effect}    
\end{figure}

\begin{figure}[h]
    \centering
    \subfigure[]{\includegraphics[width=0.49\linewidth]{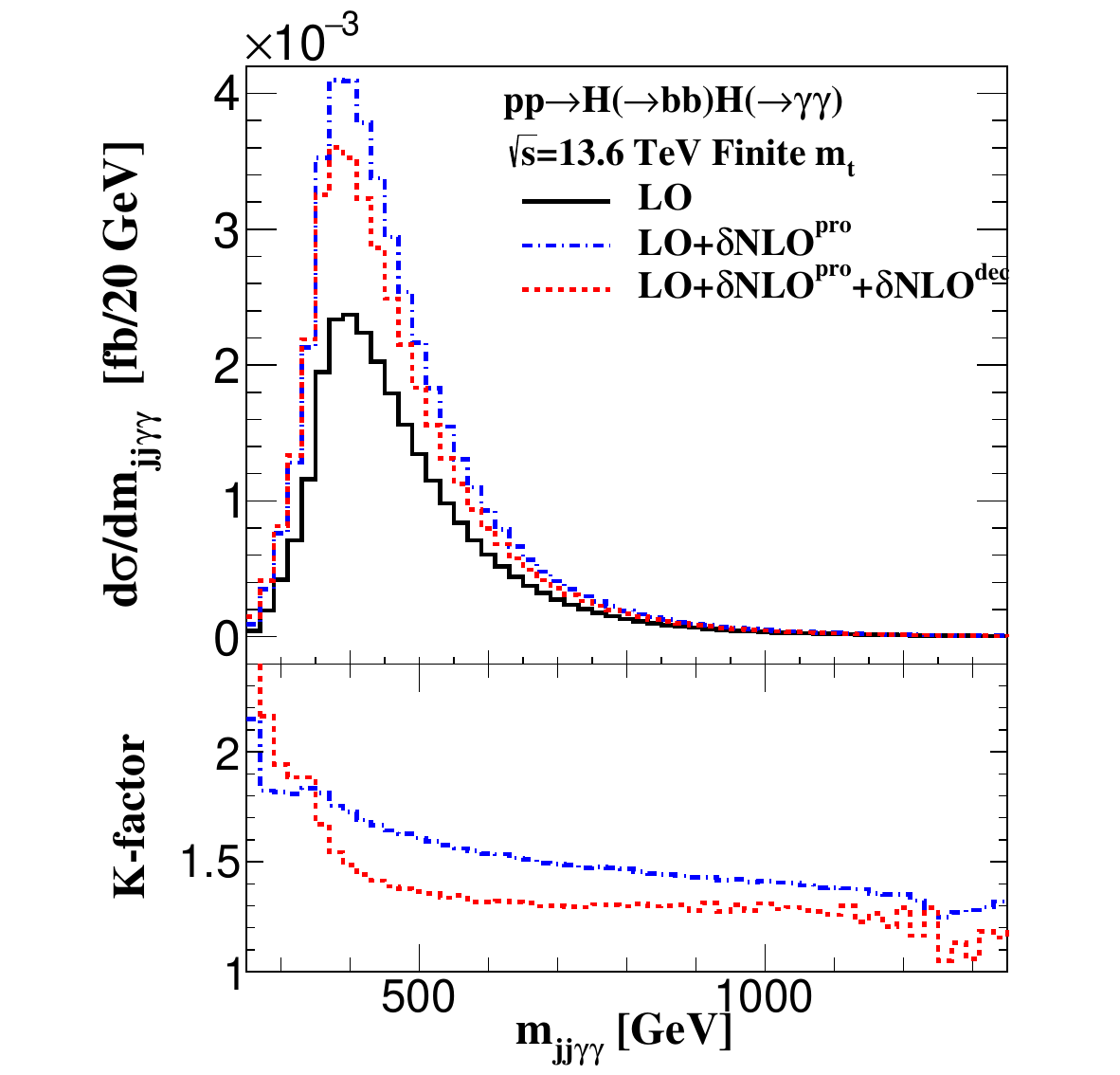}}
   \subfigure[]{\includegraphics[width=0.49\linewidth]{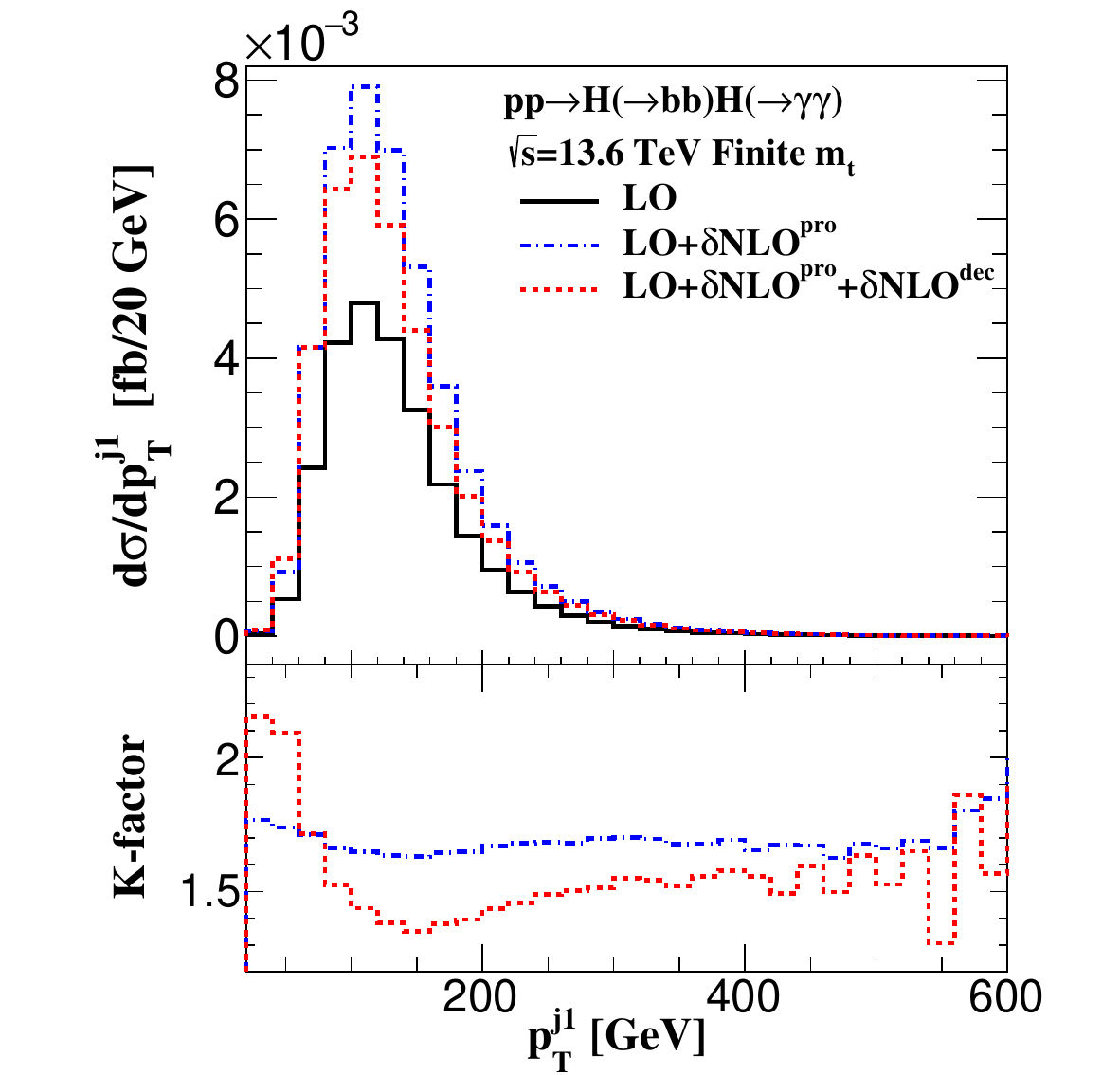}}
   \subfigure[]{\includegraphics[width=0.49\linewidth]{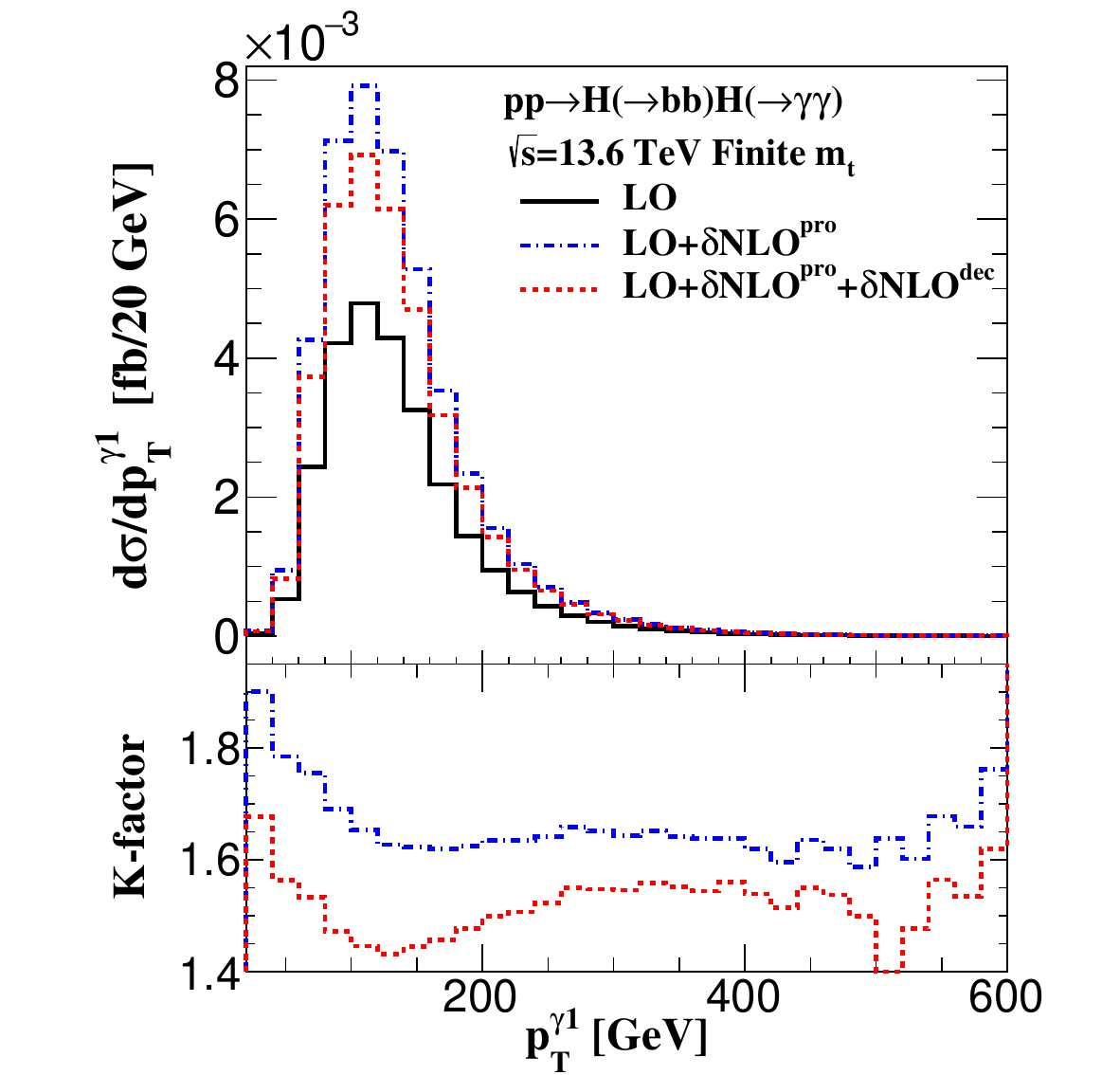}}
   \subfigure[]{\includegraphics[width=0.49\linewidth]{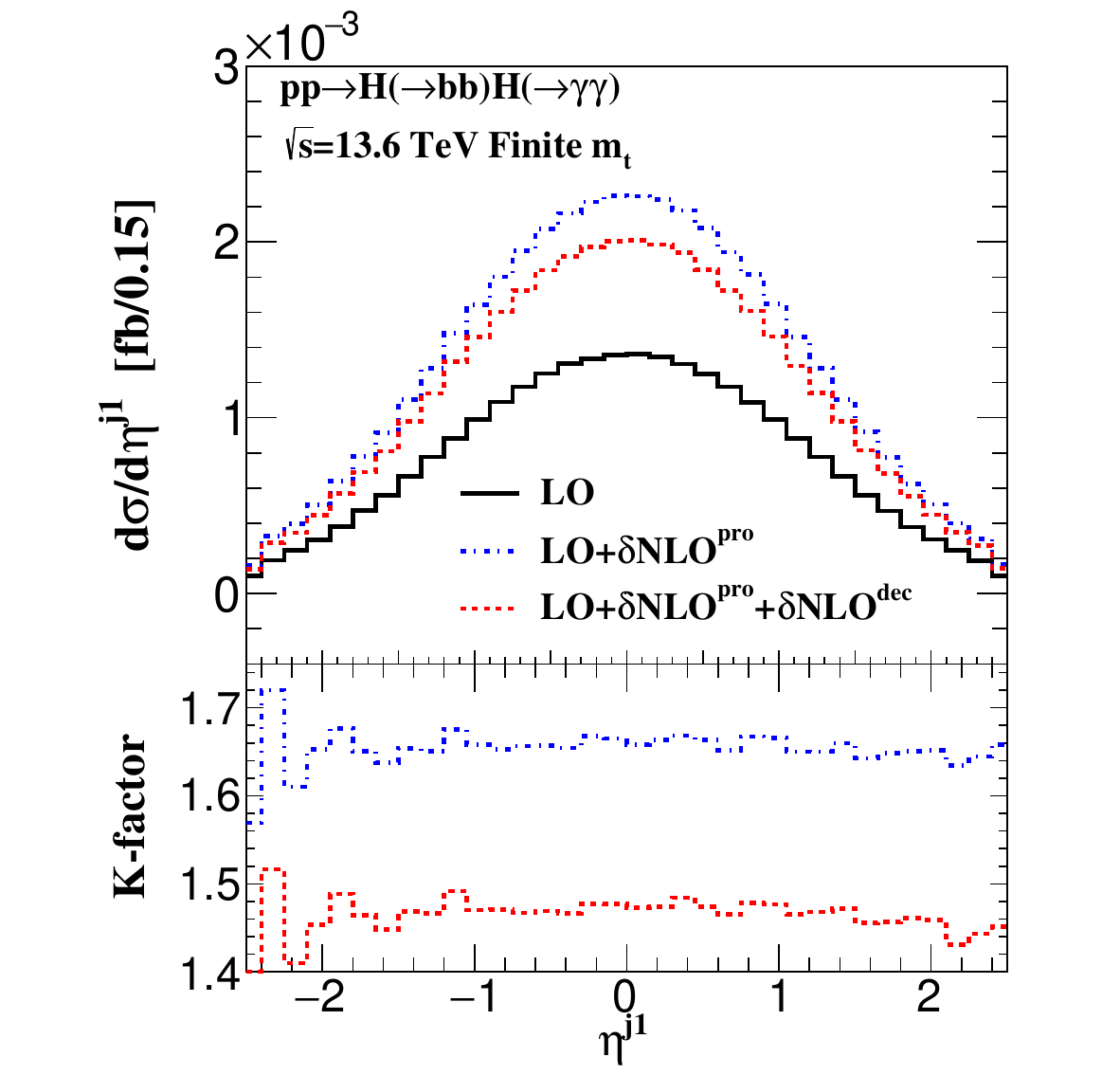}}
   \vspace{0cm}
    \caption{Kinematic distributions of the process $pp\to HH\to b\bar{b} \gamma\gamma $ after the cuts (\ref{eq:cuts}). The production cross section is calculated with finite $m_t$.
    The plot $(a)$ shows the invariant mass of the two $b$-jets and two photons.
    The plots $(b)$ and $(c)$ present the transverse momenta of the leading $b$-jet and photon, respectively. The rapidity distribution of the leading $b$-jet is shown in the plot $(d)$. The K-factor is defined as the ratio of the NLO results over the LO ones.}
    \label{fig:kinematic_NLO_effect}
\end{figure}

In figure \ref{fig:decay_NLO_effect}, we show the influence of the QCD corrections in the decay process on some kinematic distributions.
We see that the QCD corrections are most significant in the peak region of the invariant mass $m_{jj\gamma\gamma}$ distribution, which can be as large as $-25\%$.
The corrections become smaller when $m_{jj\gamma\gamma}$ moves away from the peak region.
The QCD correction to the transverse momentum of the leading $b$-jet is also prominent in the peak region, reaching $-28\%$.
The impact of the QCD correction on the transverse momentum of the subleading $b$-jet 
is important in the small $p_T$ region; it can reduce the LO cross section by $-26\%$.
The rapidity distribution receives a correction in the range from $-21\%$ to $-18\%$.

In figure \ref{fig:kinematic_NLO_effect}, we present the kinematic distributions including QCD NLO corrections both in the production and the decay processes.
The NLO corrections in the production increase the  $m_{jj\gamma\gamma}$ distribution by $54\% - 82\%$ in the region of $m_{jj\gamma\gamma} < 600$ GeV, which contributes the dominant part of the cross section.
After including the NLO corrections in the decay processes, 
the distribution is enhanced further for $m_{jj\gamma\gamma}< 350$ GeV but decreased for $m_{jj\gamma\gamma} > 350$ GeV.
A similar pattern appears in the transverse momentum distribution of the leading $b$-jet.
The NLO corrections in the production improve the LO results by a factor of $1.63-1.77$,
while the NLO corrections in the decay increase and decrease the cross section for $p_T^{j1}<80$ GeV and $p_T^{j1}>80$ GeV, respectively.
The transverse momentum spectrum of the leading photon is increased by $1.62-1.78$, considering only the QCD corrections in the production.
This improvement reduces to $1.43-1.56$ after taking the QCD corrections in the decay into account as well.
The higher-order QCD corrections to the rapidity distribution of the leading $b$-jet are stable except for the region with very large rapidity.

\section{Conclusion}
\label{sec:conclusion}

The Higgs boson pair production is an important process at the LHC that can be used to measure the Higgs boson self-couplings.
A precise theoretical prediction on its cross section enables an accurate and reliable constraint extracted from the present and future data.
Though there has been great progress in the calculations of higher-order QCD or electroweak corrections to the production processes,
the QCD corrections in the Higgs boson decays are rarely included.
In this work, we present a full calculation of the QCD NLO corrections to the process $pp\to HH \to b\bar{b}\gamma\gamma$.
The importance of including the decay is reflected by the difference between the kinematic distributions of the intermediate 
and the reconstructed Higgs bosons. 
After adopting loose kinematic cuts, we have found that the QCD corrections in the decay decrease the LO result by $19\%$.
This impact is much more significant than the ${\rm N^3LO}$ QCD corrections in the production.
We also show the QCD corrections to various kinematic distributions, 
such as the invariant mass of the $b$-jets and photons, the transverse momentum and rapidity distributions of the $b$-jets,
and find that the QCD corrections in the decay process are significant, and do not change the kinematic distributions simply by a constant multiplier.

\acknowledgments

This work was supported in part by the National Science Foundation of China under grant Nos. 12275156, 12321005, 12375076 and the Taishan Scholar Foundation of Shandong province (tsqn201909011).

\appendix

\section{$H\to b\bar{b}$ decay}
\label{sec:Hbb}

The Higgs boson to bottom quarks decay 
 is the dominant channel.
The QCD ${\rm NLO}$ corrections were  calculated forty years ago \cite{Braaten:1980yq,Sakai:1980fa,Janot:1989jf,Drees:1990dq}.
The electroweak NLO corrections were obtained around the same time \cite{Fleischer:1980ub,Bardin:1990zj,Dabelstein:1991ky,Kniehl:1991ze}.
When the final-state bottom quark is taken to be massless, the inclusive and differential cross sections have been computed up to ${\rm N}^4$LO \cite{Gorishnii:1990zu,Chetyrkin:1996sr,Baikov:2005rw,Herzog:2017dtz,Davies:2017xsp,Chen:2023fba} and ${\rm N}^3$LO \cite{Anastasiou:2011qx,DelDuca:2015zqa,Mondini:2019gid}, respectively.
The differential cross sections with massive bottom quarks have been calculated at NNLO in QCD \cite{Bernreuther:2018ynm,Behring:2019oci,Somogyi:2020mmk},
and the NNLO analytical inclusive result was studied in \cite{Chetyrkin:1995pd,Harlander:1997xa,Wang:2023xud}.
In addition, the combined QCD and electroweak corrections have been computed in \cite{Mihaila:2015lwa}, and the parton shower effect is considered in \cite{Hu:2021rkt,Bizon:2019tfo}.

In our calculation, we consider massless bottom quarks but keep the Yukawa coupling $y_b$ finite.
The LO squared amplitude is given by
\begin{align}
   \left|\mathcal{M}_{\rm tree}(H\rightarrow b \bar{b})\right|^2 =2 \sqrt{2} N_c m_H^2  m_b^2(\mu) G_F,
\end{align}
where we have replaced $y_b$ by $2^{3/4} m_b(\mu)G_F^{1/2}$.
Setting $m_H=125~{\rm GeV}, N_c=3, m_b(m_H) = 2.8 ~{\rm GeV}, G_F = 1.1663787\times10^{-5} ~{\rm GeV^{-2}}$ and performing phase space integration,  
we obtain $\Gamma_{H\rightarrow bb}^{\rm LO}=1.93 ~{\rm MeV}$. 
The NLO QCD corrections to $H\to b\bar{b}$ include contributions from both real and virtual gluon effects.
For the real corrections, we need the tree-level amplitude squared of the process $H(p)\rightarrow b(p_1)+\bar{b}(p_2)+g(p_3)$,
\begin{align}
   \left|\mathcal{M}_{\rm real}(H\rightarrow b \bar{b} g)\right|^2=\frac{8 \pi   \alpha _s C_F \left(s_{12}^2+m_H^4\right)}{{s_{13}} {s_{23}}{m_H^2}}\left|\mathcal{M}_{\rm tree}\right|^2 
\end{align}
with $s_{ij} \equiv (p_i+p_j)^2$.
In the soft or collinear limit of the gluon's momentum, i.e., $s_{13}\to 0$ or $s_{23}\to 0$, the integration of this amplitude squared becomes divergent.
In numerical calculation, we subtract these divergences using dipole counter-terms which have the same infrared structure but are easy to integrate analytically.

The virtual loop corrections in this process are easy to compute, 
and the results contain both ultraviolet and infrared divergences, which are regularized by working in $d$-dimensional space-time with $d=4-2\epsilon$.
Adopting the $\overline{\rm MS}$ and on-shell schemes for the Yukawa coupling and the bottom quark wave function, respectively,
the contribution from the counter-terms is given by
\begin{align}
 \mathcal{M}_{\rm CT}=-\frac{3\alpha_s C_F}{4\pi\epsilon} \mathcal{M}_{\rm tree} \, ,
\end{align}
which would cancel the ultraviolet divergences in the loop corrections.
The sum of the loop corrections and counter-terms is   
\begin{align} & 2{\rm Re} \left[\mathcal{M}_{\rm loop+CT}\cdot \mathcal{M}_{\rm tree}^* \right] \nn \\
 =& \frac{\as C_F}{2\pi}\frac{e^{\gamma_E \epsilon}}{\Gamma(1-\epsilon)}\left[ -\frac{2}{\epsilon^2}-\frac{1}{\epsilon}\left(3+2\log\frac{\mu_{\rm R}^2}{m_H^2}\right)-\log^2\left( \frac{\mu_{\rm R}^2}{m_H^2}\right)+\pi^2-2\right] \left| \mathcal{M}_{\rm tree}\right|^2\,.
\end{align}
The remaining infrared divergences cancel with the integrated dipole subtraction terms.
At the end, the finite result of the virtual correction reads
\begin{align}
   2{\rm Re} \left[\mathcal{M}_{\rm loop+CT}\cdot \mathcal{M}_{\rm tree}^* \right]  +\mathbf{I}(\epsilon)\times\left| \mathcal{M}_{\rm tree}\right|^2= \frac{\alpha _s C_F  }{2 \pi } \left(3\log\frac{ \mu_{\rm R}^2 }{m_H^2}+8\right) \left| \mathcal{M}_{\rm tree}\right|^2\,,
\label{eq:hbbvI}
\end{align}
where the operator $\mathbf{I}(\epsilon)$ is the integrated dipole subtraction term  \cite{Catani:1996vz,Gleisberg:2007md}. After performing the phase space integration, our result agrees with those in \cite{DelDuca:2015zqa,Gorishnii:1990zu,Gorishnii:1991zr,Baikov:2005rw}.

\section{$H\rightarrow \gamma\gamma$ decay}
\label{sec:Haa}

The Higgs boson to a photon pair decay is a clean channel in discovering the Higgs boson.
The QCD NLO corrections to this decay channel have been calculated in the heavy top quark limit \cite{Zheng:1990qa,Djouadi:1990aj,Dawson:1992cy}
and with full quark mass dependence \cite{Melnikov:1993tj,Djouadi:1993ji,Fleischer:2004vb,Harlander:2005rq,Aglietti:2006tp}.
Higher-order results are also available \cite{Maierhofer:2012vv,Davies:2021zbx,Niggetiedt:2020sbf}.
Since the NLO QCD corrections do not contribute to the final result, as discussed around eq.(\ref{eq:dhrr}),
we consider the Higgs boson decay to a photon pair at LO.
The relevant one-loop amplitude squared is given by  \cite{Spira:1995rr} 
\begin{align}
   \left|\mathcal{M}_{\varA{one-loop}}(H\rightarrow \gamma\gamma)\right|^2 
   =\frac{G_F \alpha^2 m_H^4}{4\sqrt{2} \pi ^2}\left| \sum_f N_c Q_f^2 A_f(\tau_f)+A_W(\tau_W)\right|^2,
\end{align}
where $N_c$ is the color factor, $Q_f$ is the electric charge of the fermion $f$, and the variables $\tau_{f,W}$ are defined by
\begin{align}
    \tau_f=\frac{m_H^2}{4m_f^2}\,, \qquad   \tau_W=\frac{m_H^2}{4m_W^2}\,.
\end{align}
The amplitude $A_f$ and $A_W$ can be expressed as
\begin{align}
A_f(\tau) & = \frac{2[\tau+(\tau-1)f(\tau)]}{\tau^2}\,, \nn \\
A_W(\tau) & = -\frac{2\tau^2+3\tau+3(2\tau-1)f(\tau)}{\tau^2}\,,
\end{align}
where the function $f(\tau)$ is given by 
\begin{align}
   f(\tau) =  \left\{\begin{matrix}
 {\rm arcsin^2}\sqrt{\tau} &, &\quad \tau \le 1 \, , \\
-\frac{1}{4}\left[  \log \frac{1+\sqrt{1-\tau^{-1}}}{1-\sqrt{1-\tau^{-1}}}-i\pi\right]^2 &, & \quad\tau > 1 \,.
\end{matrix}\right.
\end{align}

\bibliographystyle{JHEP}
\bibliography{ref.bib}

\end{document}